\documentclass[12pt,preprint]{aastex}

\newcommand\kms{\ifmmode \mathrm{km\,s^{-1}} \else $\mathrm{km\,s^{-1}}$\fi}

\shorttitle{Photometry and radial velocities of Cepheids}
\shortauthors{Bersier}

\begin{document}


\title{Fundamental parameters of Cepheids. V. \\
Additional photometry and radial velocity data for southern Cepheids\footnote{
Based on observations obtained at the European Southern Observatory, La Silla}}


\author{D. Bersier}
\affil{Harvard-Smithsonian Center for Astrophysics, 60 Garden St,
Cambridge, MA 02138}
\email{dbersier@cfa.harvard.edu}

\begin{abstract}

I present photometric and radial velocity data for Galactic Cepheids,
most of them being in the southern hemisphere.  There are 1250 Geneva
7-color photometric measurements for 62 Cepheids, the average
uncertainty per measurement is better than $0.01^m$.  A total of 832
velocity measurements have been obtained with the CORAVEL radial
velocity spectrograph for 46 Cepheids.  The average accuracy of the
radial velocity data is 0.38 \kms. There are 33 stars with both
photometry and radial velocity data. I discuss the possible binarity
or period change that these new data reveal.  I also present
reddenings for all Cepheids with photometry.  The data are available
electronically.

\end{abstract}


\section{Introduction}

A significant amount of photometry and radial velocity data for
Cepheids had been published by \citet{ber94a} and
\citet{ber94b}. These have been used to devise a new version of the
Baade-Wesselink method \citep{bbk97}.  However most Cepheids in
Bersier, Burki \& Kurucz's sample are in the northern hemisphere. New
observations have been obtained for (mostly) long-period Cepheids
visible from the southern hemisphere.  I present here these new
observations, along with older unpublished data. The photometry is in
the Geneva 7-color system and the radial velocity data have been
obtained with the CORAVEL radial velocity scanner \citep{bmp79}.

\section{Observations}

\subsection{Photometry}

\begin{figure}
\epsscale{0.8}
\plotone{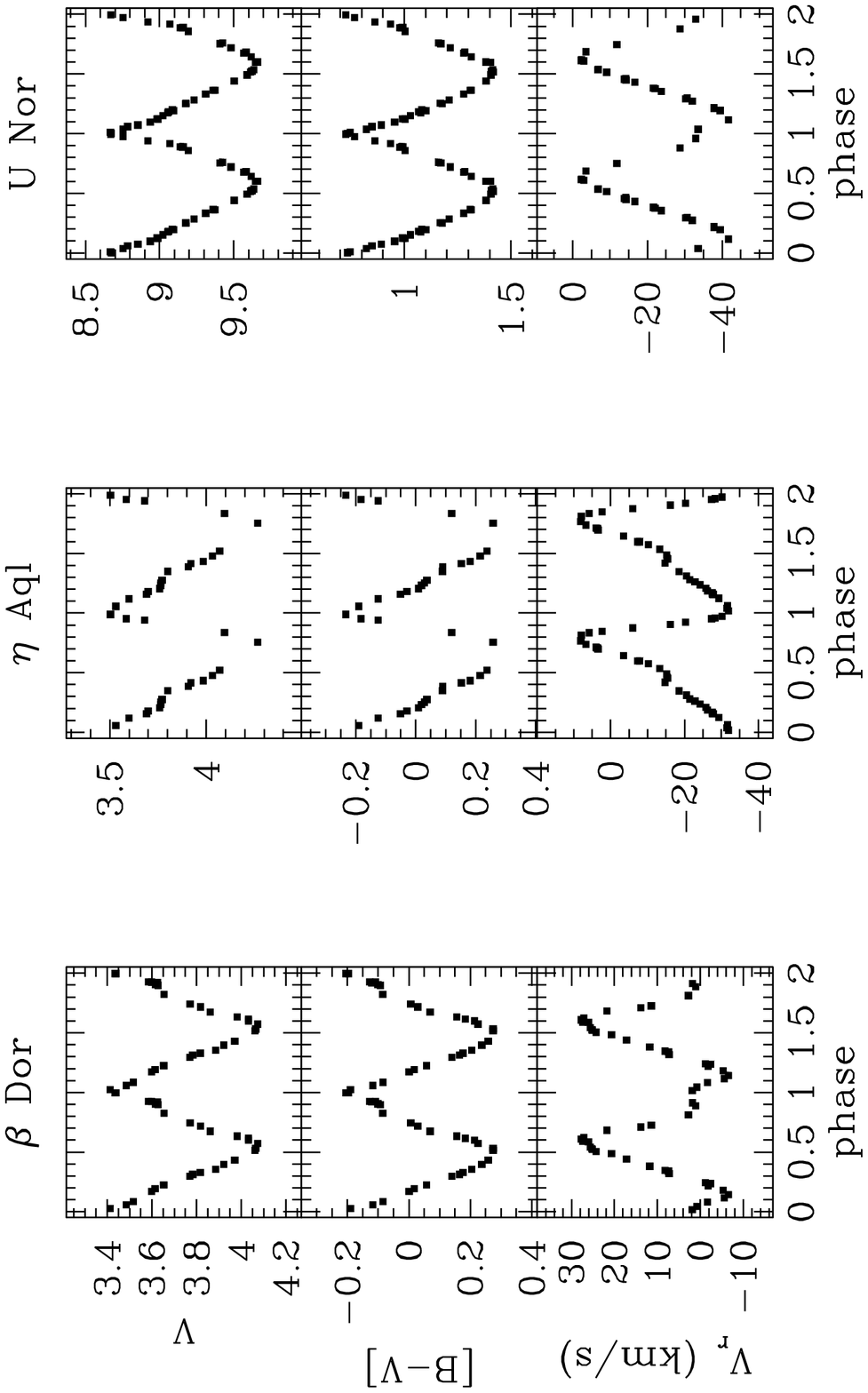}
\caption{The light- (top), color- (middle), and radial velocity (bottom)
curves for $\beta$ Dor (left), $\eta$ Aql (center) and U Nor (right).
\label{fig_lc}}
\end{figure}

A search in the Geneva photometry database revealed that several
Cepheids already had a substantial number of measurements. These data
could constitute a basis for expanding Bersier et al.'s (1997) efforts
to determine Period-Radius and Period-Luminosity relations via the
Baade-Wesselink method. I present these old unpublished
data\footnote{the present paper contains only unpublished data. Other
Cepheid data can be found in \citet{ber94a} and references therein.}
together with new data obtained in several runs during 1996 and
1997. Like the data given in \citet{ber94a}, the measurements are in
the Geneva 7-color system (Golay 1980, Rufener 1988) and most have
been obtained with the 70-cm Swiss telescope at La Silla Observatory.
The instrument used is a photometer \citep{b76} that measures each
filter several times per second; the exposure is stopped after a
minimum signal-to noise ratio has been reached in each filter; the
integration time was at least three minutes.  Given that most of our
stars are brighter than $m_V = 10$ the uncertainty is better than
$0.01^m$ for virtually all measurements. Furthermore all measurements
have been obtained in photometric conditions. Table~\ref{tbl_nph}
lists all 62 Cepheids that have data. The 1250 individual measurements
in seven colors are given in Table~\ref{tbl_ph}\footnote{All the data
presented in this paper are available electronically at \\
\url{ftp://cfa-ftp.harvard.edu/pub/dbersier/}}.  Forty-three stars
have more than 20 measurements. Figure~\ref{fig_lc} presents examples
of light and color curves for well-observed Cepheids.

\subsection{Radial velocity}

Most observations were obtained in several runs on the 1.5 meter
Danish telescope at ESO La Silla in 1996 and 1997, hence these data
are contemporaneous with most of the photometry presented above. I
used the CORAVEL spectrograph, described in detail in
\citet{bmp79}. The instrument was optimized to yield accurate radial
velocities through a cross-correlation method. The light is dispersed
and then goes through a mask (based on the spectrum of Arcturus)
before being detected by a photomultiplier. An arc spectrum is
obtained just before and just after each star exposure, to provide a
good wavelength solution.  The observing setup is such that the
cross-correlation function (CCF) is viewed in real-time. This allows
to stop the exposure when the CCF has a sufficient signal-to-noise.  A
Gaussian is fitted to the observed cross-correlation function to yield
the velocity.

A total of 46 Cepheids have been observed, representing 832
measurements.  Twenty-four stars have 15 data points or more. Some
stars have very few measurements though. They have been left out of
the target list early in the program, when it has been realized that
it would not be possible to obtain extensive phase coverage for all
Cepheids before the instrument decommissioning.  Table~\ref{tbl_nvr}
gives the name of each program Cepheid and the respective number of
measurements.  Individual measurements and their associated
uncertainties are given in Table~\ref{tbl_vr}.  Given that most of the
program stars are fairly bright, the average error per measurement is
0.38 \kms.
Examples of well-covered velocity curves are given in
Fig.~\ref{fig_lc}.

\section*{Results and discussion}

\subsection{Reddening}

\begin{figure}
\epsscale{1}
\plotone{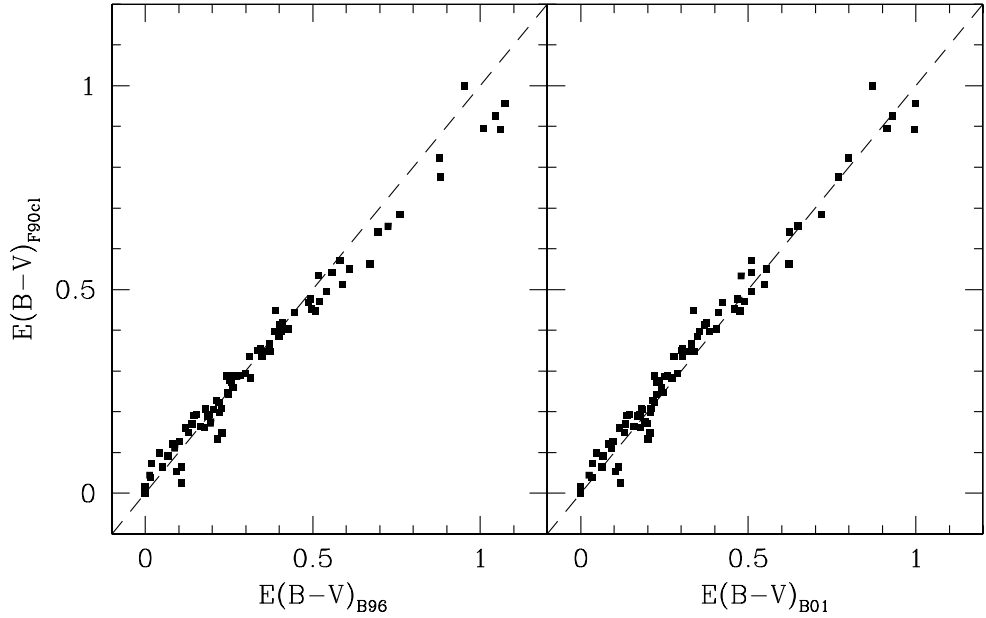}
\caption{Comparison between the reddenings of \citet{f90} and
mine. The left panel shows reddenings determined with the calibration
given in \citet{b96}.  There is clearly a scale error for $E(B-V)
\gtrsim 0.5$.  The right panel shows the color excesses computed with
the new calibration given in this paper. The line is the one-to-one
correspondence.
\label{fig_ebv}}
\end{figure}

Following the method of \citet{b96} it is possible to determine the
color excess for Cepheids. This is based on the method of \citet{f90},
using two colors that are linearly related to each other.  In
\citet{b96} there was a hint of a possible systematic difference with
Fernie's scale. This became clearly apparent with the large number of
stars available here when comparing both sources of color excesses
(see Fig.~\ref{fig_ebv}), there is a problem for stars with $E(B-V) >
0.5$. The same problem was apparent when comparing my color excesses
with those for Cepheids in open clusters \citep{fw87}, which is not
surprising given that Fernie's scale is ultimately tied to the
reddening given by \citet{fw87}. The source of this problem might be
in the small number of calibration stars with large $E(B-V)$ used in
\citet{b96}.  In order to remedy this situation, I determined a new
calibration of the color excess, using more stars with $E(B-V) > 0.5$.
As in \citet{b96} I used F- and G-type supergiants in clusters and OB
associations from the list of \citet{afp90} and supergiants with low
reddening from \citet{g91}. I also tried to use different combinations
of colors. The best results have been obtained with
\begin{equation}
E(B-V) = 0.887*(B_2-V_1) - 1.623*m_2-0.913
\end{equation}
where $m_2 = (B_1-B_2)-0.457(B_2-V_1)$.
The average error on the reddening so determined is $0.03^m$.
Figure~\ref{fig_ebv} presents a comparison of the new reddenings
with those of \citet{f90}, the agreement is now much better;
a fit gives $E(B-V)_{F90} = 0.98 E(B-V)_{here} + 0.02$.
These new color excesses for 85 stars are given in Table~\ref{tbl_nph}.

\subsection{Comments on individual stars}

In most cases, I compared the CORAVEL data with already published
radial velocity data, usually using the periods given in \citet{s89}.
This is because a number of Cepheids have been suspected to be binary
or to have a changing period.

\paragraph{U Car} It is clearly a binary. My data are offset from Coulson
\& Caldwell's (1985, hereafter CC) velocities by more than 10 \kms.

\paragraph{VY Car} The large phase offset between the CORAVEL+CC
data and \citet{s55} data suggests a period change.

\paragraph{XX Car} Even though there a few new data points, it is enough
to show that this star is a binary. The CORAVEL data are offset by 
about 30 \kms. with respect to CC's data.

\paragraph{V Cen} \citet{s89} suggested that this could be a binary. Adding data
from \citet{s55}, \citet{le80} and \citet{g81}, the evidence for binarity
is marginal. Lloyd Evans velocities are shifted by only $\sim 2$ \kms\ with
respect to CORAVEL+Gieren. Stibbs' data only add scatter.

\paragraph{V659 Cen} Photometric measurements made around $HJD = 44600$
are shifted in phase with respect to measurements obtained around $HJD= 50100$.

\paragraph{S Cru} The CORAVEL data are shifted by 2.7 \kms\ with respect to
Gieren's (1981) data. The period giving the best Fourier fit ($P =
4.689838^d$) is longer than what is predicted by \citet{s89} ephemeris
($P = 4.689392^d$).

\paragraph{T Cru} The binarity suspected by \citet{s89} is confirmed. The
data from \citet{le80} are shifted by about 4 \kms\ with respect to
mine. \citet{s55} data are also significantly shifted, giving a
lower limit on the amplitude of $\sim 8\,\kms$.

\paragraph{SY Nor} Binarity was suspected by \citet{m77}. \citet{ber94b}
confirmed the binary nature on the basis of radial velocity
measurement.  With the data published here, and adding the data of
\citet{mcs92} it is possible to constrain the orbital period at
$P_{orb} = 551.7^d$. When folded with that period, the orbital
amplitude is approximately 30 \kms.

\paragraph{AQ Pup} Even though I have only two measurements, adding the
data from \citet{s55}, \citet{cc85} and \citet{bms88} clearly shows
that the period is changing.

\paragraph{Y Sgr} \citet{s89} gives evidence that this is a binary. The present
data seem to support an orbital period around 10000 days.

\paragraph{WZ Sgr} The data from \citet{cc85} are clearly shifted with
respect to the CORAVEL data. This confirms the binary nature of this
Cepheid \citep{s89}. The measurements from \citet{go96} do not show
any shift in $\gamma$-velocity with respect to mine.

\paragraph{RZ Vel} Using data from \citet{s55}, \citet{le80} and \citet{cc85},
this star seems to show evidence for binarity and for a changing period.

\paragraph{AH Vel} This star is a known binary, although the orbital period
is not known yet (Szabados 1989, and references therein). The
$\gamma$-velocity determined from CORAVEL data is almost identical to that
given by \citet{s55}. Unfortunately the present data do not put much
constraint on the orbital period.

Several Cepheids in this sample are known binaries.  The list of known
binaries for which I present data include U Aql, $\eta$ Aql, YZ Car,
XX Cen, DX Gem, S Mus, U Sgr, X Sgr and Y Sgr.  Apart from those that
have a published orbit, evidence for binarity include the presence of
a UV-bright companion, long-term shift in the $\gamma$-velocity or
evidence from the photometry. One of the primary motivations for
publishing the present data set is that it will provide additional
radial velocities allowing to obtain orbits for several binary
Cepheids. Also, some light and velocity curves are good enough to be
used for a Baade-Wesselink analysis.

\acknowledgments

It is a pleasure to thank J.-C. Mermilliod, F. Kienzle and F. Pont for
their help with the observations, and S. Udry for extracting the
velocity data. I thank the referee for comments that improved the
presentation of this paper.
This work has been supported by NSF grant AST-9979812.
I also acknowledge partial support from the Swiss NSF through grant
8220-050332.

\clearpage


\clearpage

\clearpage




\begin{deluxetable}{l r c l r c l r c l r c}
\tabletypesize{\scriptsize}
\tablecaption{Number of photometric measurements and color excess
for each  Cepheid\label{tbl_nph}}
\tablewidth{0pt}
\tablehead{
\colhead{Star}\hfill & \colhead{$N$} & \colhead{$E(B-V)$} & %
\colhead{Star}\hfill & \colhead{$N$} & \colhead{$E(B-V)$} & %
\colhead{Star}\hfill & \colhead{$N$} & \colhead{$E(B-V)$} & %
\colhead{Star}\hfill & \colhead{$N$} & \colhead{$E(B-V)$} \\
}
\startdata
SZ Aql	                  & 23 & 0.624 & VW Cen                       &  21 & 0.337 & V508 Mon\tablenotemark{a} & 20 & 0.328 & Y Sgr                   & 15 & 0.186 \\
TT Aql	                  & 28 & 0.509 & XX Cen                       &  25 & 0.240 & R Mus                     & 20 & 0.083 & WZ Sgr                  & 29 & 0.424 \\
FF Aql                    & 6  & 0.221 & AZ Cen                       &  4  & 0.179 & S Mus                     & 24 & 0.208 & RY Sco                  & 27 & 0.770 \\
$\eta$ Aql                & 19 & 0.131 & KN Cen                       &  24 & 0.929 & UU Mus                    & 21 & 0.370 & KQ Sco                  & 31 & 0.915 \\
CO Aur\tablenotemark{a}	  & 31 & 0.214 & V659 Cen                     &  14 & 0.202 & S Nor\tablenotemark{a}    & 49 & 0.170 & Y Sct                   & 26 & 0.798 \\
SS CMa	                  & 27 & 0.554 & $\delta$ Cep                 &  3  & 0.067 & U Nor                     & 41 & 0.996 & Z Sct                   & 21 & 0.511 \\
U Car	                  & 27 & 0.273 & AV Cir                       &  4  & 0.385 & QZ Nor                    & 11 & 0.233 & RU Sct                  & 24 & 0.998 \\
VY Car	                  & 27 & 0.227 & R Cru\tablenotemark{a}       & 111 & 0.138 & V340 Nor\tablenotemark{a} & 51 & 0.303 & ST Tau\tablenotemark{a} & 17 & 0.305 \\
WZ Car	                  & 27 & 0.348 & S Cru\tablenotemark{a}       & 113 & 0.160 & Y Oph                     & 3  & 0.648 & SZ Tau\tablenotemark{a} & 24 & 0.289 \\
XX Car	                  & 24 & 0.318 & T Cru\tablenotemark{a}       & 111 & 0.147 & V440 Per\tablenotemark{a} & 35 & 0.226 & EU Tau\tablenotemark{a} & 24 & 0.198 \\
XY Car	                  & 23 & 0.375 & X Cru\tablenotemark{a}       & 111 & 0.249 & X Pup                     & 22 & 0.410 & S TrA                   & 13 & 0.046 \\
XZ Car	                  & 24 & 0.330 & SU Cru                       &  22 & 0.872 & RS Pup                    & 29 & 0.476 & $\alpha$ UMi            & 5  & 0.000 \\
YZ Car	                  & 37 & 0.353 & BG Cru                       &  24 & 0.103 & VZ Pup                    & 28 & 0.489 & RY Vel                  & 30 & 0.622 \\
AQ Car	                  & 24 & 0.114 & X Cyg\tablenotemark{a}       &  16 & 0.220 & AP Pup                    & 6  & 0.181 & RZ Vel                  & 30 & 0.278 \\
FR Car	                  & 17 & 0.301 & DT Cyg\tablenotemark{a}      &  23 & 0.033 & AQ Pup                    & 29 & 0.548 & SW Vel                  & 26 & 0.339 \\
GI Car                    & 4  & 0.191 & $\beta$ Dor                  &  29 & 0.025 & LS Pup                    & 28 & 0.469 & AH Vel                  & 33 & 0.034 \\
IT Car                    & 4  & 0.177 & DX Gem\tablenotemark{a}      &  21 & 0.460 & MY Pup                    & 29 & 0.111 & DR Vel                  & 28 & 0.717 \\
l Car	                  & 31 & 0.133 & $\zeta$ Gem\tablenotemark{a} &  27 & 0.000 & V335 Pup                  & 6  & 0.246 & T Vul\tablenotemark{a}  & 24 & 0.064 \\
SU Cas	                  & 3  & 0.258 & V473 Lyr\tablenotemark{a}    &  29 & 0.119 & S Sge                     & 5  & 0.095 & SV Vul\tablenotemark{a} & 16 & 0.511 \\
DL Cas	                  & 4  & 0.479 & T Mon                        &  32 & 0.211 & U Sgr\tablenotemark{a}    & 50 & 0.406 &                         &    &       \\
V636 Cas\tablenotemark{a} & 32 & 0.541 & BE Mon\tablenotemark{a}      &  18 & 0.516 & W Sgr\tablenotemark{a}    & 68 & 0.092 &                         &    &       \\
V Cen                     & 19 & 0.258 & V465 Mon\tablenotemark{a}    &  18 & 0.283 & X Sgr                     & 13 & 0.207 &                         &    &       \\
\enddata

\tablenotetext{a}{Data published in \citet{ber94a}. The color excess given
here supersedes that given in \citet{b96}.}

\end{deluxetable}

\clearpage


\begin{deluxetable}{l c c r c r r r r r r}
\tabletypesize{\scriptsize} \tablecaption{Photometric data
for each Cepheid\label{tbl_ph}}
\tablewidth{0pt} 
\tablehead{
\colhead{Star}\hfill & \colhead{HJD-2400000}          & \colhead{$P$\tablenotemark{a}} & 
\colhead{$V$}        & \colhead{$Q$\tablenotemark{a}} & \colhead{$U-B$}   & 
\colhead{$V-B$}      & \colhead{$B_1-B$}              & \colhead{$B_2-B$} & 
\colhead{$V_1-B$}    & \colhead{$G-B$} 
}
\startdata 
SZ Aql & 49528.825 & 3 & 8.946 & 3 & 2.554 & -1.042 & 1.425 & 1.110 & -0.211 & -0.211 \\
SZ Aql & 49529.767 & 1 & 8.894 & 3 & 2.461 & -0.954 & 1.382 & 1.126 & -0.127 & -0.113 \\
SZ Aql & 49534.778 & 3 & 8.283 & 3 & 2.449 & -0.708 & 1.293 & 1.164 &  0.099 &  0.176 \\
SZ Aql & 49537.708 & 3 & 8.574 & 3 & 2.601 & -1.026 & 1.441 & 1.083 & -0.194 & -0.178 \\
SZ Aql & 49539.720 & 4 & 8.807 & 4 & 2.772 & -1.179 & 1.511 & 1.060 & -0.341 & -0.348 \\
SZ Aql & 49541.716 & 4 & 9.037 & 4 & 2.876 & -1.286 & 1.563 & 1.037 & -0.440 & -0.475 \\
SZ Aql & 49543.677 & 3 & 9.159 & 3 & 2.865 & -1.293 & 1.543 & 1.027 & -0.448 & -0.482 \\
SZ Aql & 49548.682 & 2 & 8.060 & 3 & 2.367 & -0.408 & 1.154 & 1.255 &  0.388 &  0.511 \\
SZ Aql & 49549.694 & 2 & 8.040 & 2 & 2.374 & -0.452 & 1.187 & 1.238 &  0.342 &  0.454 \\
SZ Aql & 49558.650 & 3 & 9.018 & 3 & 2.859 & -1.278 & 1.564 & 1.031 & -0.434 & -0.467 \\
\enddata
 
\tablenotetext{a}{$P$ is the weight of the $V$ magnitude, it goes from
4 (excellent) to 0 (bad data); $Q$ is the weight for the colors; see
\citet{r88} for further details.}

\tablecomments{Table~\ref{tbl_ph} is presented in its complete form in
the electronic version of the journal. Only a fraction is shown here
for guidance regarding its form and content.}

\end{deluxetable}

\clearpage



\begin{deluxetable}{l r l r l r l r l r l r l r l r }
\tabletypesize{\scriptsize}
\tablecaption{Number of velocity measurements per Cepheid\label{tbl_nvr}}
\tablewidth{0pt}
\tablehead{
\colhead{Star}\hfill & \colhead{$N$} & \colhead{Star}\hfill & \colhead{$N$} & 
\colhead{Star}\hfill & \colhead{$N$} & \colhead{Star}\hfill & \colhead{$N$} &
\colhead{Star}\hfill & \colhead{$N$} & \colhead{Star}\hfill & \colhead{$N$}
}
\startdata
U Aql 	   & 38 & XY Car\tablenotemark{a} & 6  & V659 Cen                & 3  & %
R Mus  & 3  & AQ Pup                   & 2  & RU Sct 		      & 33 \\
SZ Aql	   & 31 & XZ Car\tablenotemark{a} & 6  & R Cru                   & 35 & %
S Mus  & 3  & LS Pup                   & 1  & S TrA  		      & 3  \\
TT Aql	   & 43 & YZ Car\tablenotemark{a} & 14 & S Cru                   & 40 & %
UU Mus & 23 & U Sgr\tablenotemark{b,c} & 32 & RY Vel                  & 8  \\
$\eta$ Aql & 28 & AQ Car\tablenotemark{a} & 15 & T Cru                   & 39 & %
U Nor  & 25 & X Sgr                    & 2  & RZ Vel                  & 18 \\
U Car	   & 16 & l Car                   & 19 & X Cru                   & 34 & %
SY Nor & 29 & Y Sgr                    & 17 & SW Vel                  & 24 \\
VY Car     & 16 & V Cen                   & 21 & SU Cru\tablenotemark{a} & 31 & %
X Pup  & 10 & WZ Sgr                   & 35 & AH Vel\tablenotemark{a} & 4  \\
WZ Car     & 6  & VW Cen                  & 1  & $\beta$ Dor             & 30 & %
RS Pup & 14 & Y Sct                    & 9  &                         &    \\
XX Car     & 6  & XX Cen                  & 34 & DX Gem\tablenotemark{b} & 33 & %
VZ Pup & 2  & Z Sct                    & 8  &                         &    \\
\enddata

\tablenotetext{a}{More CORAVEL data published by \citet{pbm94}}
\tablenotetext{b}{More CORAVEL data published by \citet{ber94a}}
\tablenotetext{c}{More CORAVEL data published by \citet{mmb87}}

\end{deluxetable}

\clearpage


\begin{deluxetable}{l c r r}
\tabletypesize{\scriptsize} \tablecaption{Radial velocity measurements
for each Cepheid\label{tbl_vr}}
\tablewidth{0pt} 
\tablehead{
\colhead{Star}\hfill & \colhead{HJD-2400000} & \colhead{$V_r\ (\kms)$} &
\colhead{$\sigma_{V_r}\ (\kms)$} } 
\startdata 
U Aql &    48076.734 &  18.21 & 0.37 \\
U Aql &    48808.691 & -12.64 & 0.38 \\
U Aql &    50224.595 &  -3.44 & 0.34 \\
U Aql &    50226.563 &  10.20 & 0.48 \\
U Aql &    50227.506 & -22.84 & 0.51 \\
U Aql &    50227.598 & -22.96 & 0.46 \\
U Aql &    50230.586 &  -7.57 & 0.40 \\
U Aql &    50274.800 &  11.62 & 0.30 \\
U Aql &    50275.773 &   8.94 & 0.35 \\
U Aql &    50276.783 & -21.80 & 0.32 \\
\enddata

\tablecomments{Table~\ref{tbl_vr} is presented in its complete form in
the electronic version of the journal. Only a fraction is shown here
for guidance regarding its form and content.}

\end{deluxetable}

\end{document}